\documentclass[12pt]{article}
\usepackage{graphics}
\usepackage{amssymb}
\usepackage{amsmath}

\newcommand{\rf}[1]{(\ref{#1})}
\newcommand{\beq}{\begin{equation}}
\newcommand{\eeq}{\end{equation}}
\newcommand{\bea}{\begin{eqnarray}}
\newcommand{\eea}{\end{eqnarray}}

\newcommand{\e}{\mbox{e}}
\renewcommand{\d}{\mbox{d}}
\newcommand{\g}{\gamma}

\newcommand{\lam}{\lambda}

\renewcommand{\b}{\beta}
\renewcommand{\a}{\alpha}
\newcommand{\n}{\nu}
\newcommand{\m}{\mu}

\newcommand{\oh}{\frac{1}{2}}

\newcommand{\tr}{\mathrm{tr}}
\newcommand{\ra}{\rangle}
\newcommand{\rra}{\right\rangle}
\newcommand{\la}{\langle}
\newcommand{\lla}{\left\langle}

\newcommand{\cD}{{\cal D}}

\newcommand{\cO}{{\cal O}}

\newcommand{\tW}{{\tilde{W}}}

\newcommand{\tZ}{{\tilde{Z}}}

\newcommand{\tw}{{\tilde{w}}}

\newcommand{\sla}{\sqrt{\lam}}

\begin{document}

\begin{center}

{ \large \bf Summing over all Topologies in CDT String Field Theory}

\vspace{30pt}

{\sl J.\ Ambj\o rn}$\,^{a,b}$, {\sl R.\ Loll}$\,^{b}$,
 {\sl W.\ Westra}$\,^{c}$ and
{\sl S.\ Zohren}$\,^{d}$

\vspace{24pt}

{\footnotesize

$^a$~The Niels Bohr Institute, Copenhagen University\\
Blegdamsvej 17, DK-2100 Copenhagen \O , Denmark.\\
{ email: ambjorn@nbi.dk}\\

\vspace{10pt}

$^b$~Institute for Theoretical Physics, Utrecht University, \\
Leuvenlaan 4, NL-3584 CE Utrecht, The Netherlands.\\
{ email: r.loll@uu.nl}\\

\vspace{10pt}

$^c$~Department of Mathematics, University of Iceland,\\
Dunhaga 3, 107 Reykjavik, Iceland\\
{ email: wwestra@raunvis.hi.is}\\

\vspace{10pt}

$^d$~Mathematical Institute, Leiden University,\\
Niels Bohrweg 1, 2333 CA Leiden, The Netherlands\\
{email: zohren@math.leidenuniv.nl}\\

}
\vspace{48pt}

\end{center}


\begin{center}
{\bf Abstract}
\end{center}
By explicitly allowing for topology to change as a function of time, two-dimensional quantum 
gravity defined through causal dynamical triangulations gives rise to a new continuum string
field theory. Within a matrix-model formulation we show that -- rather remarkably -- the 
associated sum over all genera can
be performed in closed form, leading to {\it a} nonperturbative definition of CDT string field theory.
We also obtain explicit formulas for the 
$n$-loop correlation functions. Our construction exhibits interesting parallels with previous, purely
Euclidean treatments.

\vspace{12pt}
\noindent


\newpage

\section{Introduction}\label{sec1}

By general acknowledgement, string theory is in need of a genuinely nonperturbative 
definition if it is to serve as a fundamental theory of all interactions.
String theory's perturbative expansion does not serve this purpose, since it 
is only asymptotic and becomes use\-less when the string coupling approaches order one.
It is thus of great interest to study toy models like non-critical 
string theory for which one can define the sum over 
all genera, a possibility first outlined in the seminal papers
\cite{sum1,sum2,sum3}. Unfortunately, the results of such summations are plagued
by ambiguities: even in the simplest cases, like pure two-dimensional
Euclidean quantum gravity ($c=0$ non-critical string theory) and all 
unitary non-critical string theories, the 
``nonperturbative'' partition function is not Borel-summable
\cite{silvestrov,moore,david-x,zinn} (see also
\cite{marino} for a recent lucid discussion).
Attempts to perform the summation in these theories typically result in 
complex partition functions or, perhaps more accurately, constructions which {\it do}
obtain a real partition function are often guided by the desire for a
real outcome, rather than by applying stringent physical criteria.
(For other approaches see, for example, \cite{stochastic}). 

Given this somewhat unsatisfactory state of affairs, any new physical model for which one
has analytic control over the sum over topologies is potentially
valuable, to understand possible ambiguities in the summation and their genericity, 
to explore their physical 
meaning, and to see whether new insights can be gained into the nature of   
surfaces of infinite genus, a topic which has maybe not received
the attention it deserves.

The present article deals with just such a new system in two spacetime dimensions.
Its origin lies in the attempt to formulate a nonperturbative theory of four-dimensional quantum 
gravity based on conventional quantum field theory, more precisely, the 
path integral approach, in the framework of ``Causal Dynamical
Triangulations'' (or ``CDT" for short), which has led to a variety of intriguing and 
unprecedented results \cite{higher-d,higher-d1}. However, the construction can in
principle be applied in any dimension, leading to nonperturbative models of
dynamical quantum geometry with causal, Lorentzian properties. The case of two
spacetime dimensions is a particularly attractive testing ground because of the
availability of a whole range of analytical tools. The original, strictly causal and purely gravitational model
was solved exactly in 2d \cite{al}, and has since been generalized by the inclusion of
matter \cite{2dmatter,2dmatter1} and isolated (cap or branching) points where causality is 
violated. The latter has culminated in the recent definition of a fully fledged CDT string field
theory (in zero-dimensional target space), which is the subject of the remainder of
the present work.

In the mid-nineties, Kawai, Ishibashi and collaborators developed a string field theory for
non-critical strings \cite{sft,stf1,sft2,sft3,sft4,sft5}. 
While its original starting point was the explicit realization of non-critical
string theory in terms of dynamical triangulations (of the Euclidean variety), it eventually
was formulated entirely in the language of continuum field theory. In 
\cite{cdt-sft,cdt-sft1} we repeated 
this continuum analysis for the two-dimensional 
CDT model of quantum gravity. We demonstrated how it is connected to 
the conventional matrix model of 2d Euclidean quantum 
gravity \cite{cdt-matrix,cdt-matrix1} 
and how the {\it continuum} Dyson-Schwinger
equations of the CDT string field theory are related to a particular matrix 
model.\footnote{Since no scaling limit is needed in this matrix model, the situation
is similar to what happens in topological 2d gravity
where the Kontsevich matrix model directly describes the continuum
gravity theory \cite{marshakov,marshakov1,dijkgraaf} 
(see \cite{rastelli} for a recent 
description in terms of D-branes). As will become clear below, this similarity is not accidental.}
In what follows, we will use this matrix-model formulation to perform the sum over all topologies 
in the continuum CDT string field theory and compute a number of associated geometric observables,
so-called loop-loop correlation functions.
It turns out that this can be done in a surprisingly simple manner.

The remainder of this article is organized as follows: in section \ref{sec2}
we introduce the CDT matrix model and related observables,
in section \ref{sec3} we show how to calculate these quantities summed
over all genera, in section \ref{sec4} we put our results into the context of
previous Euclidean models, and finally, 
in section \ref{sec5} we discuss the nature of our nonperturbative results.

\section{The CDT matrix model}\label{sec2}

The approach of causal dynamical triangulations
aims to define quantum gravity nonperturbatively via a path integral over 
geometries $[g_{\m\n}]$,
\beq\label{2.1}
Z(G_{\rm N},\lam)= \int \cD [g_{\m\n}] \; \e^{-S[g_{\m\n}]},
\eeq
where $S$ denotes the (Euclidean) Einstein-Hilbert action of gravity, $G_{\rm N}$ is Newton's constant
and $\lambda$ the cosmological constant. We are presently interested
in the case of two spacetime dimensions for which the action is given by
\beq\label{2.2}
S[g_{\m\n}] = -\frac{1}{2\pi G_{\rm N}} \int d^2\xi\ \sqrt{\det{g_{\m\n}}}\;R 
+ \lam \int d^2\xi\  \sqrt{\det g_{\m\n}}.
\eeq
One proceeds in two steps, first setting up the CDT lattice regularization of the path integral in 
{\it Lorentzian} signature, and then rotating each individual triangulation in the Lorentzian
ensemble to Euclidean signature.
The resulting Euclidean path integral \rf{2.1}
differs from a standard Euclidean path integral since by construction each 
geometric ``path" has a time-foliation and thus a ``memory" of the causal properties it possessed
before the Wick rotation. We refer the reader to the original 
article \cite{al} for technical details.

The observables we will discuss in this article 
are generalized Hartle-Hawking amplitudes involving 2d geometries with a fixed number of 
boundaries.
The simplest one is that for a single boundary, with all boundary points sharing a common
time (in the time-foliation referred to above). This so-called Hartle-Hawking or disc amplitude  
$W(l)$ is the amplitude that the spatial boundary (consisting of a closed, one-dimensional loop) 
has length $l$. Depending on whether one thinks of the boundary as being initial or final,
it is the amplitude of the universe of length $l$ vanishing into or coming from ``nothing".
A generalized such amplitude is one where the spatial boundary (still at equal time) consists of $n$ 
disconnected loop components with specified lengths
$l_1,\ldots,l_n$. Spacetime
may remain disconnected or reconnect in various ways
at other times.  We denote the corresponding amplitudes $W_d(l_1,\ldots,l_n)$, where 
the subscript ``d'' indicates that for $n>1$ it includes the 
possibility of having disconnected universes. The 
connected component of $W_d(l_1,\ldots,l_n)$ will be called $W_c(l_1,\ldots,l_n)$.
To lowest order $n\! =\! 1$ we have $W_d(l)=W_c(l)=W(l)$, while for $n\! =\! 2$
\beq\label{2.x}
W_c(l_1,l_2)= W_d(l_1,l_2)-W(l_1)W(l_2),
\eeq
which generalizes to higher-loop correlators in standard fashion.

Instead of considering a situation where the boundary lengths $l_i$
are kept fixed, one can introduce by hand boundary
cosmological constants $x_i$ and include corresponding terms $x_i\cdot l_i$ 
in the action. Now the lengths $l_i$ are allowed to fluctuate, 
but their average values are determined by the values 
of the $x_i$. This leads to another kind of  generalized Hartle-Hawking
wave functions, denoted by $\tW_d(x_1,\ldots,x_n)$ and
 $\tW_c(x_1,\ldots,x_n)$, and 
related to the previous $W_{c,d}(l_1,\ldots,l_n)$ by Laplace transformation according to
\beq\label{2.2a}
\tW_{c,d}(x_1,\ldots,x_n) = \int_0^\infty \d l_1\cdots \d l_n \;
\e^{-(x_1l_1+\cdots+x_nl_n)}\;W_{c,d}(l_1,\ldots,l_n).
\eeq 

The Dyson-Schwinger equations, an infinite set of 
coupled equations relating  $\tW_c(x_1,\ldots,x_n)$ for different $n$,
follow from the continuum CDT string field theory as 
developed in \cite{cdt-sft}.
They are obtained from the cubic matrix model model
\beq\label{2.3}
Z_N(g,\lam)= \int \d M \; 
\exp \left[ -\frac{N}{g}\, \tr \left( \lam M - \frac{1}{3}
\; M^3 \right) \right],
\eeq
where $M$ is an $N\times N$ Hermitian matrix.
In this matrix model the observables introduced above take the form
\bea\label{2.4}
\tW_d(x_1,\ldots,x_n) &=& \frac{1}{N^n}
\lla \tr \left(\frac{1}{x_1-M}\right) \cdots\; \tr\! 
\left(\frac{1}{x_n-M}\right) 
\rra,\\
\label{2.4a}
\tW_c(x_1,\ldots,x_n) &=& 
\frac{1}{N^n}\lla \tr \left(\frac{1}{x_1-M}\right) 
\cdots \tr\left(\frac{1}{x_n-M}\right) 
\rra_{{\rm connected}},
\eea
where the expectation value of an operator $\cO$ is defined by 
\beq\label{2.5}
\la \cO(M)\ra = \frac{1}{Z_N(g,\lam)} 
\int \d M \;\cO(M)\; 
\exp \left[ -\frac{N}{g}\, \tr \left( \lam M - \frac{1}{3}\; M^3\right)\right].
\eeq

Of course an expression like \rf{2.3} is formal and should
always be understood as a suitable expansion in powers of $M$.
This is most clearly exhibited by making the variable change 
\beq\label{2.6}
M= {\sqrt{t \lam}}\;Y -\sqrt{\lam},~~~~t= 
\frac{g}{\lam^{3/2}}\equiv {\rm e}^{-1/G_{\rm N}},
\eeq  
where $t$ is a dimensionless coupling constant, which is related to the
coupling $g$ of eq.\ (\ref{2.3}) and the gravitational coupling $G_{\rm N}$ as
specified. In terms of the matrix $Y$ and the coupling $t$ the matrix integral
reads
\beq\label{2.7}
Z_N(t(g,\lam),\lam) = \exp\left(\frac{2 N}{3t}\right)\; (t\lam)^{N^2/2}\; 
\int \d Y \; 
\exp\left[ -N \, \tr \left(Y^2 -\frac{\sqrt{t}}{3}Y^3\right) \right],
\eeq
which after expanding $\exp (\sqrt{t} Y^3/3)$ and performing the Gaussian integrals
becomes a formal power expansion in $t$.
The coefficient of $t^n$ in this power series
is positive, in agreement with 
the interpretation that it represents the number of 2d 
surfaces with a time foliation with $n$ degenerate points
(see \cite{cdt-sft} for details).  
Let us emphasize that the pre-factor, which does not have a power expansion in $t$, will 
cancel in any expectation values of observables and should not really be considered
part of the partition function. In fact we could have defined the partition function as 
\beq\label{2.8}
\tZ_N (t) =  \frac{\int \d Y \; 
\exp\left[ -N \, \tr \left(Y^2 -\frac{\sqrt{t}}{3}Y^3\right) \right]}{\int \d Y \; 
\exp\left[ -N \, \tr\ Y^2  \right]},
\eeq
which would then only contain positive powers of $t$. However, we find it convenient
to use \rf{2.7} in the rest of this paper, with the understanding that the exponential 
growth of $Z_N(t)$ for $t \to 0$ should not be considered as a reflection of unphysical 
behaviour, but is 
merely a consequence of a particular choice of normalization. 

\section{Summing over all genera}\label{sec3}

In reference \cite{cdt-matrix} the matrix model (\ref{2.3}) was related to the CDT Dyson-Schwinger equations 
by (i) introducing into the latter an expansion parameter $\a$, which kept track 
of the genus of the two-dimensional spacetime, and (ii) identifying this 
parameter with $1/N^2$, where $N$ is the size of the matrix in the matrix integral.
The $1/N$-expansion of our matrix model therefore plays 
a role similar to the $1/N$-expansion originally introduced 
by 't Hooft \cite{thooft}: it reorganizes an asymptotic expansion in a coupling constant
($t$ in our case) into convergent sub-summations in which the $k$th summand appears with a 
coefficient $N^{-2k}$. In QCD applications, the physically relevant value is $N=3$, 
to which the leading-order terms in the large $N$-expansion can under favourable circumstances
give a reasonable approximation. 

As we will see, for the purposes of solving our string 
field-theoretic model
nonperturbatively, an additional expansion in inverse powers of $N$ (and thus an identification
of the contributions at each particular genus) is neither essential nor does it provide any new insights. 
This means that we will consider the entire sum over topologies ``in one go", which simply amounts
to setting $N=1$,  upon which
the matrix integral \rf{2.3} reduces to the ordinary integral\footnote{Starting 
from a matrix integral for $N\times N$-matrices 
like (\ref{2.3}), performing a {\it formal} expansion
in (matrix) powers commutes with setting $N=1$, as follows from the following property of
expectation values of products of traces, which holds for
any $n=1,2,3,\dots$ and any set of non-negative
integers $\{ n_k\} $, $k=1,\dots , 2n$, such that $\sum_{k=1}^{2n}n_k =2n$. For any particular choice of
such numbers, consider
\beq\label{foot1}
\langle \prod_{k=1}^{2n} \Big(\frac{1}{N}\tr M^{n_k}\Big) \rangle \equiv
\frac{\int \d M \, \e^{-\oh \tr M^2} 
\prod_{k=1}^{2n} \Big(\tr M^{n_k}/N\Big)}{\int 
\d M \, \e^{-\oh \tr M^2}}=
\sum_{m=-n}^{n} \omega_m N^{m},
\eeq
where the last equation {\it defines} the numbers $\omega_m$ as coefficients in the power expansion in 
$N$ of the expectation value. Now, we have that
\beq\label{foot2}
\sum_{m=-n}^{n}\omega_m = (2n-1)!!
\eeq
independent of the choice of partition $\{ n_k\}$.
The number $(2n-1)!!$ simply counts the ``Wick contractions" of $x^{2n}$ which we
could have obtained directly as the expectation value $\langle x^{2n}\rangle$, evaluated
with a one-dimensional Gaussian measure.
In the model at hand, we will calculate sums of the form
$\sum_{m=-n}^{n}\omega_m$ directly, since we 
are summing over all genera {\it without} introducing an
additional coupling constant for the genus expansion. 
In other words, the dimensionless
coupling constant $t$ in this case already contains the information about the splitting
and joining of the surfaces, and the coefficient of $t^k$ 
contains contributions from 2d geometries 
whose genus ranges between 0 and $[k/2]$. We cannot 
disentangle these contributions further unless we introduce $N$ as an extra parameter.}
\beq\label{3.1}
Z(g,\lam) = \int \d m \; 
\exp \left[ -\frac{1}{g} 
\left( \lam m - \frac{1}{3}\; m^3\right)\right],
\eeq 
while the observables \rf{2.4} can be written as 
\beq\label{3.2}
\tW_d(x_1,\ldots,x_n) = \frac{1}{Z(g,\lam)} 
\int \d m\; \frac{\exp \left[ -\frac{1}{g} 
\left( \lam m - \frac{1}{3}\; m^3\right)\right]}{(x_1-m)\cdots (x_n-m)}.
\eeq
Again, these integrals should be understood as formal power series
in the dimensionless variable $t$ as mentioned below eq.\ \rf{2.7}.
Any choice of an integration contour which makes the integral well 
defined and reproduces the formal power series is a potential
nonperturbative definition of these observables. However, different
contours might produce different nonperturbative contributions
(i.e.\ which cannot be expanded in powers of 
$t$), and there may even be nonperturbative contributions 
which are not captured by any choice of integration contour. 
As usual in such situations, additional
physics input is needed to fix these contributions.

To illustrate the point, let us start by evaluating the partition function given in 
\rf{3.1}. We have to decide on an integration path in the 
complex plane in order to define the integral. One possibility is to take a 
path along the negative 
axis and then along either the positive or the negative imaginary 
axis. The corresponding integrals are 
\beq\label{3.2a}
Z(g,\lam)= \sqrt{\lam}\; t^{1/3} F_{\pm} (t^{-2/3}),~~~
F_{\pm} (t^{-2/3}) =2\pi \; e^{\pm i\pi/6}{\rm Ai}(t^{-2/3}\e^{\pm 2\pi i/3}),
\eeq
where Ai denotes the Airy function. Both $F_\pm$ 
have the same asymptotic expansion
in $t$, with positive coefficients. Had we chosen the integration path 
entirely along the imaginary axis we would have obtained ($2\pi i$ times)
${\rm Ai}(t^{-2/3})$, but this has an asymptotic expansion 
in $t$ with coefficients of oscillating sign, which is at odds with its
interpretation as a probability amplitude. In the notation of \cite{as} we have
\beq\label{3.2b}
F_{\pm}(z) = \pi \Big({\rm Bi}(z) \pm i {\rm Ai}(z)\Big),
\eeq 
from which one deduces immediately 
that the functions $F_{\pm}(t^{-2/3})$ are not real.
However, since ${\rm Bi}(t^{-2/3})$ grows like 
$e^{\frac{2}{3t}}$ for small $t$ while ${\rm Ai}(t^{-2/3})$ 
falls off like $e^{-\frac{2}{3t}}$, their imaginary parts are exponentially small 
in $1/t$ compared to the real part, and therefore do not contribute to
the asymptotic expansion in $t$.
An obvious way to {\it define} a partition function which is real and shares the
same asymptotic expansion is by symmetrization,
\beq
\oh (F_+ +F_-)\equiv \pi {\rm Bi}.
\eeq
The situation parallels the one encountered in the double scaling limit of the 
``old'' matrix model \cite{david-x}, and discussed in detail in \cite{marino},
but is less complicated. We will return to a discussion of this in the next section. 

Presently, let us collectively denote by $F(z)$ any of the functions 
$F_{\pm}(z)$ or $\pi {\rm Bi}(z)$, leading to the
tentative identification
\beq\label{3.3}
Z(g,\lam) = \sqrt{\lam}\; t^{1/3} \, F\Big(t^{-2/3}\Big),~~~~F''(z) = z F(z),
\eeq 
where we have included the differential equation satisfied by the Airy functions for
later reference. In preparation for the computation of the 
observables $\tW_d(x_1,\ldots,x_n)$ we introduce the
dimensionless variables
\beq\label{3.4}
x= y\,\sqrt{\lam},~~m= g^{1/3} \b,~~~~~
\tW_d(x_1,\ldots,x_n) = \lam^{-n/2} \tw_d(y_1,\ldots,y_n).
\eeq
Assuming $y_k > 0$, we can write
\beq\label{3.5}
\frac{1}{y-t^{1/3}\,\b}
= \int_0^{\infty} \d\a
\; \exp\left[-\left(y-t^{1/3}\b\right)\;\a\right].
\eeq
We can use this identity to re-express the pole terms 
in eq.\ \rf{3.2} to obtain the 
integral representation 
\beq\label{3.6}
\tw_d(y_1,\ldots,y_n) =  
\int_0^{\infty}\prod_{i=1}^n \d \a_i
\; \e^{-(y_1\a_1+\cdots +y_n\a_n)}\; 
\frac{F\Big(t^{-2/3}-t^{1/3}\sum_{i=1}^n\a_i\Big)}{F\Big(t^{-2/3}\Big)}
\eeq
for the amplitude with dimensionless arguments.
From the explicit expression of the Laplace transform, eq.\ \rf{2.2a}, 
we can now read off the generalized Hartle-Hawking amplitude as function of the
boundary lengths,
\beq\label{3.7}
W_d(l_1,\ldots,l_n) = 
\frac{F(t^{-2/3}-t^{1/3}\sqrt{\lam}\,(l_1+\cdots+l_n))}{F(t^{-2/3})}.
\eeq
For the special case $n=1$ we find
\beq\label{3.8}
W(l) = \frac{F(t^{-2/3}-t^{1/3}\sqrt{\lam}\,l)}{F(t^{-2/3})}
\eeq
for the disc amplitude, 
together with the remarkable relation\footnote{Formula \rf{3.8a} has a structure
quite similar to the one encountered in the ``old'' matrix model 
\cite{ajm,staudacher}, where again the multi-loop correlator is ``almost'' 
a function of $l_1+\cdots+l_n$ only.}
\beq\label{3.8a}
W_d(l_1,\ldots,l_n)=W(l_1+\cdots+l_n).
\eeq
By Laplace transformation this formula implies the relation
\beq\label{3.8c}
\tW_d(x_1,\ldots,x_n)= 
\sum_{i=1}^n \frac{\tW(x_i)}{\prod_{j\ne i}^n (x_j-x_i)}. 
\eeq
Before turning to a discussion of the nonperturbative
expression for $W(l)$ we have just derived, let us remark that the asymptotic
expansion in $t$ of course agrees with that obtained
by recursively solving the CDT Dyson-Schwinger equations.
Using the standard asymptotic expansion of the Airy function \cite{as}
one obtains
\beq\label{3.9}
W(l) = \e^{-\sqrt{\lam} \,l} \; \e^{t \, h(t,\sqrt{\lam} \,l)}\;\;
\frac{\sum_{k=0}^\infty  c_k \;t^k 
\;(1-t\sqrt{\lam}\,l)^{-\frac{3}{2} 
k-\frac{1}{4}}}{\sum_{k=0}^\infty c_k \;t^k},
\eeq
where the coefficients $c_k$ are given by $c_0=1$, $c_k=\frac{1}{k!} \left(\frac{3}{4}\right)^k
\left(\frac{1}{6}\right)_k \left(\frac{5}{6}\right)_k$, $k>0$.
In (\ref{3.9}), we have rearranged the exponential factors to exhibit the exponential
fall-off in the length variable $l$, multiplied by a term containing the function 
\beq\label{3.10}
h(t,\sqrt{\lam}\,l) = \frac{2}{3t^2} 
\left[(1-t\sqrt{\lam}\,l)^{3/2}-1+\frac{3}{2}t\sqrt{\lam}\,l\right],
\eeq
which has an expansion in positive powers of $t$.

Finally, let us derive an expression for the amplitude $\tW(x)$.
Since $F(z)$ satisfies the Airy differential equation, we have the identity
\beq\label{4.1} 
F\Big(t^{-2/3}-t^{1/3}\a\Big) =  
 \left( \frac{\d^2}{\d\a^2} +t\a\right) F\Big(t^{-2/3}-t^{1/3}\a\Big).
\eeq
Inserting this into eq.\ \rf{3.6} and performing the partial integrations,
we obtain a first-order differential equation for $\tw(y)$, namely, 
\beq\label{4.2}
\tw(y) = \left(y^2-t \frac{\d}{\d y}\right) w(y) + 
\frac{t^{1/3} F'\Big(t^{-2/3}\Big)}{F\Big(t^{-2/3}\Big)} -y.
\eeq
Its solution is given by 
\beq\label{4.3}
\tw(y) = t^{-1} \e^{-\frac{1}{t}\Big(y-\frac{1}{3}y^3\Big)}
\int_y^{\infty} \d v \; \e^{\frac{1}{t}\Big(v-\frac{1}{3}v^3\Big)}
(v-1-f(t)),
\eeq
where the function $f(t)$ is defined by
\beq\label{4.4}
f(t) = \frac{t^{1/3} F'\Big(t^{-2/3}\Big)}{F\Big(t^{-2/3}\Big)} -1.
\eeq
Performing a variable shift $v=y+t \, \xi$ in the integral in \rf{4.3}, 
$\tw (y)$ is conveniently written as
\beq\label{4.5}
\tw(y) = \int_0^{\infty} \d \xi\; \e^{-(y^2-1)\xi} 
\; e^{-t y \xi^2 -t^2 \xi^3/3} \; \Big[(y-1) -f(t) +t\xi\Big].
\eeq
From \rf{4.5} it follows by expanding the exponential containing
$t$ that $\tw(y)$ has an asymptotic
expansion in $t$ (for $y > 1$). The same expansion represents $\tw(y)$ 
by an expansion in inverse powers of $(y+1)$, corresponding 
to the expansion \rf{3.9}. Explicitly, one finds
\beq\label{4.6}
\tw(y) = \frac{1}{y+1} + t \frac{y+3}{4(y+1)^3} + O(t^2),
\eeq
which, as already stated, of course agrees with the perturbative 
expansion of $\tw(y)$ derived previously in CDT string field theory \cite{cdt-sft}.

\section{Relation with other models}\label{sec4}

We should perhaps not be too surprised to meet the Airy function as part of
our nonperturbative analysis, 
since it has already appeared previously in
non-critical string theory, more specifically, in the so-called
2d topological quantum gravity. This theory can be described in 
two ways. On the one hand, a set of observables of the theory, the intersection
indices of Riemann surfaces, can be calculated 
by the Kontsevich matrix integral, which {\it is} the matrix generalization of the Airy 
function. This is a cubic integral like \rf{2.3},
but with $\lam$ a matrix, which gives it more structure
than \rf{2.3} and allows for the calculation of 
the intersection indices. Analogous to our case
\rf{2.3}, it is a matrix representation of a {\it continuum} theory, namely,
2d topological quantum gravity. 

On the other hand, 2d topological quantum gravity
also has a ``conventional'' one-matrix representation, which in fact is the 
simplest one possible. Recall that the $(p,q)$ minimal conformal 
field theories coupled to 2d Euclidean quantum gravity in the 
cases $(p,q)=(2,2m-1)$, $m=2,3,\ldots$, can be described as 
double-scaling limits of one-matrix models with certain
fine-tuned matrix potentials of order at least $m+1$. Formally,
the case $m=1$, which corresponds to a somewhat degenerate $(2,1)$
conformal field theory with central charge $c= -2$, is then described
by a special double-scaling limit of the purely Gaussian matrix model
(see the review \cite{ginsparg}). In this double-scaling limit one obtains for
the so-called FZZT brane precisely the Airy function, see 
\cite{seiberg,rastelli,marino} for recent discussions. 

Our model is {\it not} equivalent to 2d Euclidean topological quantum gravity,
but is dual to it in a specific way. The arguments presented below
suggest that our CDT string field theory can also be identified as a continuum theory
associated with $c=-2$, but corresponding to the unconventional, ``wrong" branch of the KPZ
equation. -- Recall that for a given 
conformal field theory coupled to Euclidean 2d quantum gravity, 
we have the KPZ formula 
\beq\label{6.1}
\g_- = \frac{-(1-c) -\sqrt{(1-c)(25-c)}}{12}
\eeq
for the susceptibility (for spherical topology).
The parameter $\g_-$ corresponds to the ``right'' choice of branch of the 
quadratic KPZ equation, i.e.\ the branch which leads to a weakly
coupled Liouville theory as $c \to -\infty$. Choosing instead
the other branch, one obtains
\beq\label{6.2}
\g_+= \frac{-(1-c) +\sqrt{(1-c)(25-c)}}{12}= -\frac{\g_-}{1-\g_-}.
\eeq
The interpretation of this $\g_+$ in terms of matrix models and
geometry can be found in \cite{durhuus,adj,klebanov}, and for earlier
related work see \cite{wadia,kom}.
The simplest example is again given by $c=-2$: topological quantum gravity 
has $\g_-=-1$ whose dual is the ``wrong'' $\g_+ = 1/2$, which happens to be the 
value occurring generically in the theory of branched polymers 
(see, e.g., \cite{ad,ajt} for a discussion of why branched polymers
and baby universes are generic and even dominant in many situations).
While it is possible to describe 2d topological quantum gravity  
by a double-scaling limit of the Gaussian matrix model, the 
most natural geometric interpretation of the Gaussian matrix model
is in terms of branched polymers, in the sense that the integral
\beq\label{6.3}
\frac{ \int \d M \; \tr\, M^{2n} \;\e^{-\oh \tr M^2}}{\int \d M \; 
 \e^{-\oh \tr M^2}}
\eeq
can be thought of as the gluing of a boundary of length $n$ into a double-line branched 
polymer of length $n$. Since the branched polymers are also allowed to form closed
loops, their partition function contains a sum over topologies `en miniature', and
one can indeed define a double-scaling limit of the model.
When solving for the partition function in this limit, one obtains 
precisely our $Z(g,\lam)$ of eq.\ \rf{3.3}!\footnote{In the case of the branched polymer model,
the parameters $g$ and $\lam$ appearing in the double-scaling limit  
are related to the ``topology'' of the branched polymer (i.e.\ 
to $1/N^2$ in the matrix model) and to the length of the polymer, respectively. 
We refer to \cite{jk} for details on this work.} One should not jump to the conclusion
that the two models are identical, since the CDT string field theory has a much richer
structure of observables, with no obvious analogues in the branched-polymer
set-up. Nevertheless, it is obvious that the partition function captures the essentials
of the counting of branchings and joinings, which for the case of the CDT model
is insensitive to the fact that the geometries are genuinely extended. (The latter is obvious from
the nontrivial dependence on the boundary cosmological constants or, equivalently, the
boundary lengths.) 

The fact that the CDT string field theory shares some
properties of branched polymers is maybe less surprising in view of the fact that
the original two-dimensional CDT model without branching can be mapped
to a one-dimensional random-walk model \cite{charlotte}. The generalization
implemented by the CDT string field theory corresponds to adding ``branches" to
the random walks, resulting again in branched polymers.

\section{Discussion}\label{sec5}

The central result of this paper is the derivation of the explicit, 
nonperturbative expressions \rf{3.3} and \rf{3.7}
for the partition function $Z(g,\lam)$ and the Hartle-Hawking amplitude 
$W_d(l_1,\dots ,l_n)$ of the CDT string field theory, both incorporating the infamous ``sum over
topologies" (2d spacetimes of all possible genera). It is rather remarkable that
these sums can be performed and -- with hindsight -- in a manner which is technically 
not very involved. As was already the case for the two-dimensional CDT quantum gravity 
theory with fixed spacetime topology, the results are genuinely different from those of
the corresponding purely Euclidean models. Nevertheless, as we have tried to argue
above, from the point of view of conformal field theory, the CDT string field theory can 
probably be understood as the continuum theory ``in the wrong branch" 
with conformal charge $c=-2$ and susceptibility $\g_+=1/2$, and thus as ``dual" to topological
quantum gravity in two dimensions.

The nonperturbative aspects of our theory suffer from the same fundamental ambiguity 
as string theory, and which are rooted in the non-Borel summability of the perturbation series.
Beyond the perturbative expansion, which unfortunately is only
asymptotic, there is no real definition of the theory.
We have managed to sum the asymptotic series
and produce a closed formula for $W(l)$, but
like in non-critical string theory with $c \geq 0$, i.e.\ unitary 
field theories coupled to 2d Euclidean quantum gravity, we lack a 
clear physical principle which would allow us to decide which nonperturbative
completion of the perturbative expansion to choose. 

If one insists on a {\it real} partition function (in the Euclidean sector), it is natural to take 
$F(z) =\pi {\rm Bi}(z)$ in formulas 
\rf{3.3} and \rf{3.7}. However, this choice is only unique within the
matrix-model realization. 
While it is true that matrix models in non-critical string theory have been able to
incorporate physics they were not originally designed to incorporate, like the
physics of $ZZ$-branes, we are not aware of any argument that would identify
matrix models as {\it the} correct, nonperturbative definitions of continuum theories
including a sum of genera, should they indeed exist. 

Our string-field theoretic model highlights in a particularly simple and 
transparent manner the limited amount of information contained in the
perturbative expansion. When comparing explicitly the closed-form
nonperturbative results \rf{3.3} and \rf{3.8} with their asymptotic expansions,
one finds that the latter are only
good approximations in a small range of their arguments $t$ and $l$. This 
can simply be traced to the fact that the asymptotic expansion of for instance
${\rm Bi}(t^{-2/3})$, terminated after $k$ terms, is only valid for $t \ll 4/k$.
In a similar vein, 
many aspects of the nonperturbative solution could not possibly have been guessed
from the perturbative series. Consider, for instance, the behaviour of the 
nonperturbative disk amplitude
$W(l)$ as a function of the boundary length $l$ for fixed $g$ and $\lam$.
Each term in the perturbative expansion of $W(l)$ falls off exponentially as $\e^{-\sla l}$
for $\sla l \gg 1$ {\it and} is positive. However, while the full nonperturbative function $W(l)$ 
will initially decrease with increasing $l$, as expected from each of the 
(positive) terms in its asymptotic expansion, it  becomes oscillatory when $l > 1/(t\sla) = \lam/g$.  
The same oscillatory behaviour occurs  when $t$ is increased while $l$ and 
$\lam$ are 
kept fixed, i.e.\ when the coupling $g$ is increased.
This oscillation is a genuinely nonperturbative effect, which
is opposite to the behaviour of {\it each} term in the perturbative expansion
of $W(l)$ in powers of $t$ and may be indicative of a phase transition to spacetimes
completely dominated by topology changes.

\section*{Acknowledgment}
JA, RL and WW acknowledge the support by
ENRAGE (European Network on
Random Geometry), a Marie Curie Research Training Network in the
European Community's Sixth Framework Programme, network contract
MRTN-CT-2004-005616. RL acknowledges
support by the Netherlands
Organisation for Scientific Research (NWO) under their VICI
program.


\end{document}